\documentclass[sigconf]{acmart}

\usepackage{algorithm}
\usepackage{algorithmic}
\usepackage{fancyvrb}
\usepackage{pifont}  
\usepackage{graphicx}   
\usepackage{amsmath} 
\usepackage{textcomp}  
\usepackage{fancybox}    
\usepackage{float}   
\usepackage{enumitem}
\usepackage{framed}

\newcommand{\boxedcell}[1]{%
  \fbox{\strut #1}%
}

\newenvironment{myboxedtable}
{%
  \begingroup
  \setlength{\tabcolsep}{0pt}%
}
{%
  \endgroup
}

\newcommand{\tool}{\textsc{Crucio}}
\newcommand{\sotak}{\textsc{Kedavra}}
\newcommand{\sotaa}{\textsc{Arvada}}
\newcommand{\sotat}{\textsc{Treevada}}

\newcommand\blfootnote[1]{%
  \begingroup
  \renewcommand\thefootnote{}\footnote{#1}%
  \addtocounter{footnote}{-1}%
  \endgroup
}

\newcommand{\Function}[2]{\STATE \textbf{function} #1(#2)\begin{ALC@g}}
\newcommand{\EndFunction}{\end{ALC@g}\STATE \textbf{end function}}
\newcommand{\Return}[1]{\textbf{return }#1}
\newcommand{\Call}[2]{\textsc{#1}(#2)}
\newcommand{\Comment}[1]{\hfill{\textcolor{gray!70}{\footnotesize// #1}}}

\newcommand{\cmark}{\ding{51}}  % 对号
\newcommand{\xmark}{\ding{55}}  % 错号

\AtBeginDocument{%
  }

\setcopyright{acmlicensed}
\copyrightyear{2026}
\acmYear{2026}
\acmDOI{10.1145/3744916.3773147}

\acmConference[ICSE '26]{48th IEEE/ACM International Conference on Software Engineering}{April 12-18 2026}{Rio de Janeiro, Brazil}
\acmISBN{978-1-4503-XXXX-X/18/06}

\begin{document}

\title{Context-Free Grammar Inference for Complex Programming Languages in Black Box Settings}

\author{Feifei Li\textsuperscript{1,6}, Xiao Chen\textsuperscript{2}, Xiaoyu Sun\textsuperscript{3}, Xi Xiao\textsuperscript{1,6*}, Shaohua Wang\textsuperscript{4},  Yong Ding\textsuperscript{5*}, Sheng Wen\textsuperscript{5}, Qing Li\textsuperscript{6}} 
\affiliation{%
  \institution{\textsuperscript{1}Tsinghua Shenzhen International Graduate School, China}
  \country{}
}
% \email{lff23@mails.tsinghua.edu.cn,xiaox@sz.tsinghua.edu.cn}
% \email{}
\affiliation{%
  \institution{\textsuperscript{2}The University of Newcastle, Australia}
  \country{}
}
% \email{xiao.chen@newcastle.edu.au}

\affiliation{%
  \institution{\textsuperscript{3}The Australian National University, Australia}
  \country{}
}
% \email{xiaoyu.sun@anu.edu.au}

\affiliation{%
  \institution{\textsuperscript{4}Central University of Finance and Economics, China}
  \country{}
}
% School of Computer Science & lnformation Security, Guilin University of Electronic Technology, Guilin, Gusngxi, China
\affiliation{%
  \institution{\textsuperscript{5}School of Computer Science, Guangdong University of Science and Technology, Dongguan 523668, China}
  \country{}
}
\affiliation{%
  \institution{\textsuperscript{6}Peng Cheng Laboratory, China}
  \country{}
}
% \email{}

\renewcommand{\shortauthors}{Li et al.}

\begin{abstract}

\blfootnote{*Corresponding authors: Xi Xiao and Yong Ding.\\\raggedright Emails: Feifei Li (lff23@mails.tsinghua.edu.cn), Xiao Chen (xiao.chen@newcastle.edu.au), Xiaoyu Sun (xiaoyu.sun1@anu.edu.au), Xi Xiao (xiaox@sz.tsinghua.edu.cn), Shaohua Wang (davidshwang@ieee.org), Yong Ding (dingyong@gdust.edu.cn), Sheng Wen (wensheng@gdust.edu.cn), Qing Li (liq@pcl.ac.cn).}
Grammar inference for complex programming languages remains a significant challenge, as existing approaches fail to scale to real-world datasets within practical time constraints. In our experiments, none of the state-of-the-art tools, including \sotaa{}, \sotat{} and \sotak{} were able to infer grammars for complex languages such as C, C++, and Java within 48 hours.
\sotaa{} and \sotat{} perform grammar inference directly on full-length input examples, which proves inefficient for large files commonly found in such languages. While \sotak{} introduces data decomposition to create shorter examples for grammar inference, its lexical analysis still relies on the original inputs. Additionally, its strict no-overgeneralization constraint limits the construction of complex grammars.

To overcome these limitations, we propose \tool{}, which builds a decomposition forest to extract short examples for lexical and grammar inference via a distributional matrix.
Experimental results show that \tool{} is the only method capable of successfully inferring grammars for complex programming languages (where the number of nonterminals is up to 23$\times$ greater than in prior benchmarks) within reasonable time limits. On the prior simple benchmark, \tool{} achieves an average recall improvement of 1.37$\times$ and 1.19$\times$ over \sotat{} and \sotak{}, respectively, and improves F1 scores by 1.21$\times$ and 1.13$\times$. 

\end{abstract}





\maketitle
\section{Introduction}
Inferring context-free grammars (CFGs)~\cite{cremers1975} from example strings has a wide range of applications in software engineering, including code understanding~\cite{oda2015learning}, reverse engineering~\cite{moonen2001generating}, protocol specification recovery~\cite{caballero2007polyglot,narayan2015survey}, and input generation for fuzz testing~\cite{gopinath2019building,aschermann2019nautilus,alsaeed2023finding,godefroid2008grammar,nguyen2020mofuzz,srivastava2021gramatron,wang2019superion}.

Most existing grammar inference techniques operate in grey-box or white-box settings, where the internal structure or source code of the parser is accessible~\cite{blazytko2019grimoire,lin2008deriving,hoeschele2016mining}. In contrast, black-box grammar inference aims to recover the underlying grammar of a parser with no access to its internal implementation, relying solely on input samples and binary feedback (accept or reject) from the parser. The objective is to construct a grammar that approximates the oracle’s behavior, accurately predicting acceptance for arbitrary inputs.

Recent work in black-box grammar inference includes methods such as \sotaa{}~\cite{kulkarni2021learning}, \sotat{}~\cite{arefin2024fast}, and \sotak{}~\cite{li2024incremental}. While these approaches show promise on small, synthetic grammars, they struggle with real-world programming languages. 
The datasets on which \sotaa{}, \sotat{}, and \sotak{} were evaluated typically correspond to parser grammars with fewer than 20 nonterminals.
To evaluate scalability, we introduce challenging benchmarks derived from C, C++, Java, Lua, and Rust, including grammars with 20–200 nonterminals and input samples averaging over 2,000 characters. Existing tools fail to produce valid grammars on these  datasets.

Inferring grammars for realistic programming languages presents a number of challenges. First, the large size of input programs — often containing thousands of tokens — makes direct grammar inference computationally expensive. \sotaa{} and \sotat{} both generalize grammar rules by selecting bubbles to merge based on contextual similarity on original input programs, resulting in grammars that are overly complex, and inefficient to infer. Although \sotak{} introduces data decomposition to improve scalability, its lexical inference still operates on the original, undecomposed samples, resulting in a significant performance bottleneck. Furthermore, its decomposition strategy does not guarantee that the decomposed fragments preserve the grammar coverage of the original inputs, leading to potential generalization loss. \sotak{} also enforces strict constraints to prevent overgeneralization, but its reliance on sampling-based validation means that errors can go undetected. These undetected errors accumulate over time and can render future grammar generalization steps ineffective or unstable. 

To address these limitations, we propose \tool{}, a novel black-box grammar inference framework designed to scale to complex programming languages while maintaining precision and interpretability. \tool{} introduces several key innovations. First, it introduces the concept of the \textit{decomposition forest}, which systematically explores structured decompositions of input samples. This enables the discovery of short, generalizable samples while preserving grammar coverage — overcoming the generalization loss observed in \sotak{}. Second, \tool{} improves lexical inference by operating on shorter decomposed samples. It uses decision trees to classify tokens and applies the L* algorithm~\cite{angluin1987learning} to generalize token-level lexical rules efficiently and robustly. Third, for grammar inference, \tool{} constructs a distributional matrix and applies grammar generalization over parse trees while enforcing \emph{first-order swap correctness}. This ensures bounded correctness guarantees without requiring additional oracle queries beyond the initial stage.

To the best of our knowledge, \tool{} is the \textbf{first approach capable of inferring usable grammars for complex programming languages} such as C, C++, and Java in black-box settings within a practical timeframe.

This paper makes the following contributions:
\begin{itemize}[leftmargin=*]
\item We present \tool{}, a novel grammar inference framework that improves the generalization capability and scalability of black-box grammar inference. The tool is open-sourced and available at \url{https://github.com/Sinpersrect/crucio}.

\item \tool{} is the first tool capable of inferring accurate grammars for real-world programming languages within a practical time limit. On the complex benchmarks (including C, C++, Java, etc.), it is the only method that produces results within the timeout threshold. On the simple benchmarks (used in evaluating state-of-the-art (SOTA) tools and involving languages such as Lisp, JSON, Turtle etc.), \tool{} achieves performance comparable to that of the state-of-the-art grammar inference tools.

\item We propose \emph{Swap Precision}, a new grammar evaluation metric that addresses key limitations of sample-based precision. 
Swap Precision avoids biases in sample-based precision that favor invalid grammars focusing on few samples and grammars generating short samples, enabling more reliable evaluation.
\end{itemize}

\section{Background and Related Works}

\textbf{Context-Free Grammar} (CFG)~\cite{cremers1975} is a well-established mathematical formalism used to define the syntactic structure of formal languages, particularly in the specification of programming language syntax. Formally, a CFG is defined as a quadruple \( G = (V, \Sigma, R, S) \), where \( V \) denotes a finite set of nonterminal symbols, \( \Sigma \) denotes the set of terminal symbols (disjoint from \( V \)), \( R \) is a finite set of production rules of the form \( A \rightarrow \alpha \) with \( A \in V \) and \( \alpha \in (V \cup \Sigma)^* \), and \( S \in V \) is the designated start symbol.

CFGs allow the recursive definition of potentially infinite languages through a finite set of rules. For example, consider the grammar with
$V = \{S\}, \quad \Sigma = \{a, b\}, \quad \text{and production rules} \quad S \rightarrow aSb \mid ab.$
This grammar generates the language \(\{a^n b^n \mid n \geq 1\}\), which includes strings such as ``ab'', ``aabb'', ``aaabbb'', and so forth. The derivation process begins with the start symbol \( S \) and applies production rules recursively until a string consisting solely of terminal symbols is obtained.

CFGs strike a balance between expressiveness and tractability. They are powerful enough to describe the syntax of most programming languages while remaining constrained enough to support efficient parsing algorithms, such as LL, LR, and CYK parsers. Accordingly, CFGs form the syntactic foundation for compilers and interpreters of programming languages such as C++~\cite{cpp}, Java~\cite{java}, and JavaScript~\cite{javascript}. Additionally, CFGs are extensively used in validating the syntax of structured data formats like JSON~\cite{smith2015} and XML~\cite{marchal2002}, as well as in natural language processing.

\textbf{Black-box Grammar Inference.}
Black-box grammar inference was initiated by \textsc{Glade}\cite{bastani2017synthesizing}, which laid the groundwork but showed limitations when processing highly recursive grammars.
To overcome these issues, \sotaa{}\cite{kulkarni2021learning} proposed a method that generalizes grammars by merging Bubbles, enabling more effective handling of recursive structures.
Following this, \sotat{}\cite{arefin2024fast} builds on \sotaa{} by optimizing the handling of parentheses in the inferred grammar, enabling recursive application of learned rules, and further supporting the generation of deterministic grammars.
\sotaa{} and \sotat{} both infer grammars directly from the original samples, and their performance degrades as input complexity increases, making them unsuitable for large and complex inputs that may span thousands of characters. Moreover, their bubble merging strategy relies on context similarity, without taking grammar generalizability into account, which leads to limited generalization capability.
\sotak{}\cite{li2024incremental} introduced a data decomposition strategy that segments input strings into smaller units and applies incremental inference, significantly improving grammar precision, recall, and interpretability.
However, the lexical inference in \sotak{} is still performed on the original samples. Moreover, its generalization requires zero errors, a condition that becomes increasingly difficult to satisfy as the grammar complexity increases. Therefore, it does not work well on more complex grammars.
Separately, \textsc{REINAM}\cite{wu2019reinam} focuses on probabilistic context-free grammars (PCFGs) by first extracting a CFG with \textsc{Glade} and then refining it via reinforcement learning to adjust production probabilities.
More recently, Jia et al.\cite{jia2024vstar} presented \textsc{V-Star}, which uses active learning with visibly pushdown automata to accurately infer input languages with nested hierarchical structures such as XML and JSON.
However, \textsc{V-Star} tends to produce overly complex grammars and suffers from low efficiency. On relatively simple datasets, it performs significantly slower than \sotaa{} and \sotat{}.

\textbf{Grey-box Grammar Inference.}
Grey-box grammar inference operates in scenarios where full access to the oracle’s source code is unavailable, but partial internal information, such as coverage, can be utilized. \textsc{GRIMOIRE}~\cite{blazytko2019grimoire} adopts this approach by extracting grammar-like structures based on feedback from code coverage during fuzzing. While it does not construct an explicit CFG, it leverages behavioral signals to guide input generation more effectively.

\textbf{White-box Grammar Inference.} 
White-box grammar inference refers to inferring input grammars with access to the parser’s source code. Representative methods include Lin~\cite{lin2008deriving}, which employs dynamic analysis, and \textsc{AUTOGRAM}\cite{hoeschele2016mining}, which uses dynamic taint analysis to track input flow and reconstruct context-free grammars. More recently, Leon\cite{bettscheider2025symbolic} proposed a symbolic analysis approach specifically designed for recursive descent parsers, extracting grammar rules by directly analyzing the control flow and source structure. These methods all require access to the parser’s source code; when the source code is unavailable, they no longer work.

\textbf{Deep Learning.} 
While deep learning models such as RNNs have been applied to grammar inference, studies have shown that they struggle to capture the structural properties of context-free grammars~\cite{sennhauser2018evaluating,bernardy2018can,yu2019learning}. Yellin~\cite{yellin2021synthesizing} attempted to overcome this limitation by synthesizing CFGs from RNNs. However, the lack of \textit{active learning}~\cite{angluin1981}, a core mechanism in \sotaa{}, \sotat{}, and \tool{}, leads to significantly weaker results compared to these approaches.


\section{Approach}

\begin{figure*}
\setlength{\abovecaptionskip}{2pt} 
\setlength{\belowcaptionskip}{2pt} 
    \centering
    \includegraphics[width=1\linewidth]{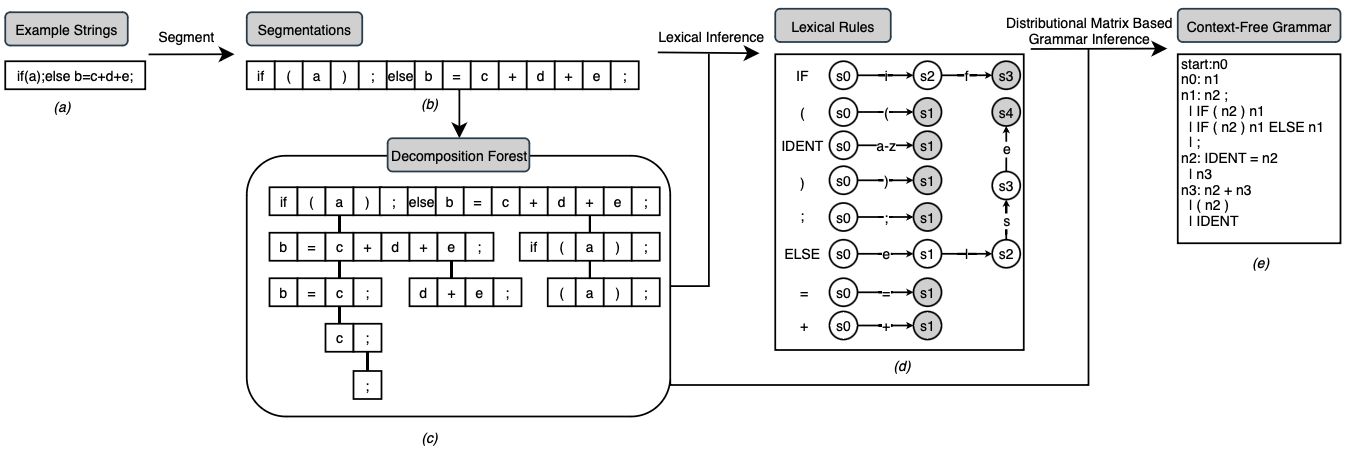}
    \caption{Workflow of \tool{}}
    \label{fig:workflow}
\end{figure*}

Figure \ref{fig:workflow} illustrates the workflow of our proposed system, \tool{}, which is specifically designed for grammar inference in complex programming languages. The overall process consists of three main stages: decomposition forest construction, lexical inference, and distributional matrix based grammar inference.

\textbf{Decomposition forest} is the core structure of \tool{}. It can construct the input sequence as a tree, and by searching for samples that meet the given conditions on the tree, it can obtain short samples that meet the given conditions for lexical or grammar inference, significantly reducing the complexity of the overall inference task. It can solve the problem of generalization loss in \sotak{}.

\textbf{Lexical Inference} obtain short samples from a decomposition forest for inference, using the L* algorithm to construct an automaton as lexical rules.

\textbf{Distributional matrix based grammar inference} begins by constructing a distributional matrix from the examples obtained from the decomposition forest, capturing the contextual distribution of the examples. Then, by repeatedly merging bubble maximal cliques, it generalizes the structure of flat parse trees, ultimately deriving a context-free grammar that satisfies first-order swap correctness. 
After constructing the grammar, \tool{} continues to search for samples in the decomposition forest that can generalize the current grammar, then reconstructs the grammar.

In the following sections, we will provide a detailed explanation of \tool{}’s core stages and the technical details behind them.

\subsection{Decomposition Forest}

We propose a novel data structure called the decomposition forest, which is composed of multiple decomposition trees. 
% Decomposition forest is a tree-structured representation
A decomposition tree is a hierarchical representation in which the children of each node correspond to samples of lower complexity than their parent.
The decomposition forest is used to search for short samples that satisfy given conditions, which can then be used for lexical or grammar inference. Simpler input samples lead to more efficient inference.
Figure \ref{fig:workflow}(b) shows an example of a decomposition forest.

Each decomposition tree is dynamically expanded. It is neither necessary nor practical to compute the entire tree upfront, as doing so would incur significant computational overhead. 
Instead, child nodes are generated on demand based on specific decomposition strategies. Currently, we have designed three decomposition strategies to compute the children of each node: 
\textbf{Binary maximum subsequence deletion}: The sequence is divided into two parts, and we alternately apply maximum subsequence deletion to either the first or the second part. %This approach is illustrated in Figure~\ref{fig:decompose_strategy}(a).
\textbf{Maximum subsequence deletion}: This strategy prioritizes the deletion of large subsequences, progressing from larger to smaller segments, with particular focus on eliminating loop structures. %This approach is illustrated in Figure~\ref{fig:decompose_strategy}(b).
\textbf{Subsequence replacement}: Replace a subsequence with a subsequence of itself, primarily to simplify or remove recursive constructs. %An illustration is provided in Figure~\ref{fig:decompose_strategy}(c).

The search algorithm traverses each tree to determine whether its root node satisfies the given condition. If the condition is met, it attempts to traverse its child nodes (at which point the decomposition tree is dynamically expanded). If no child nodes satisfy the condition, we return the value of the current node. If a child node satisfies the condition, we recursively search on that child node.

\subsection{Lexical Inference}

The purpose of lexical inference is to convert raw input strings into tokens, enabling grammar inference to be conducted at the token level. This significantly reduces the computational cost of subsequent inference processes.
In many programming languages, inputs can be highly complex and contain thousands of tokens.  
In such cases, the default lexical inference used in \sotak{} becomes inefficient, as it is performed directly on raw input samples.
To address this scalability challenge, we introduce a dedicated lexical inference framework designed for large and structurally rich inputs.

Our lexical inference procedure consists of four main steps:
\begin{itemize}[leftmargin=*]
\item \textbf{Segmentation:} Split input examples into individual tokens based on a set of predefined lexical rules.
\item \textbf{Classification:} Group the segmented tokens into lexical classes based on their contextual distribution. Tokens with the same distribution are assigned to the same class.
\item \textbf{Character-level Generalization:} For each token class, infer a finite automaton using the L algorithm to generalize at the character level.  The inferred automata serve as lexical rules for token recognition.
\item \textbf{Lexical Validation:} Validate the inferred lexical rules on the original input examples, and iteratively refine the rules when discrepancies or errors are detected.
\end{itemize}

This framework leverages \emph{decomposition forest} to search for the shortest example that enables lexical inference.  

\subsubsection{Segmentation}

\textbf{Segmentation transforms character-level analysis into token-level analysis, reducing sequence length and improving efficiency.}
The process begins by applying  a set of predefined lexical rules to segment each input string into candidate tokens.
To further refine token boundaries, we identify characters that the grammar is insensitive to, meaning those that can appear freely at token boundaries and whose repetition does not affect syntactic validity.
Such characters are excluded from the grammar.  
Among them, we specifically check whether the space character (ASCII 32) is grammar-insensitive.  
If it is, we leverage its common role as a delimiter in programming languages to refine token boundaries.
Specifically, we iteratively insert a space character at each candidate boundary and query the oracle to check whether the modified input remains valid.  
If the insertion is accepted, the boundary is confirmed; otherwise, the adjacent segments are merged.

\textbf{Example:}  
Consider the tinyc input shown in Figure~\ref{fig:workflow}(a).

\begin{myboxedtable}

\begin{tabular}{|*{20}{c|}} % 20列，根据你字符数调整
\hline
\boxedcell{\texttt{i}} & \boxedcell{\texttt{f}} & \boxedcell{\texttt{(}} & \boxedcell{\texttt{a}} & \boxedcell{\texttt{)}} & \boxedcell{\texttt{;}} & 
\boxedcell{\texttt{e}} & \boxedcell{\texttt{l}} & \boxedcell{\texttt{s}} & \boxedcell{\texttt{e}} & \boxedcell{\texttt{\textvisiblespace}} & \boxedcell{\texttt{b}} & 
\boxedcell{\texttt{=}} & \boxedcell{\texttt{c}} & \boxedcell{\texttt{+}} & \boxedcell{\texttt{d}} & \boxedcell{\texttt{+}} & \boxedcell{\texttt{e}} & \boxedcell{\texttt{;}} \\
\hline
\end{tabular}
\end{myboxedtable}

\vspace{1em}

The initial segmentation (based on predefined rules) yields the following sequence of candidate tokens:

\begin{myboxedtable}
\begin{tabular}{*{15}{c}} % 15列
\boxedcell{\texttt{if}} & \boxedcell{\texttt{(}} & \boxedcell{\texttt{a}} & \boxedcell{\texttt{)}} & \boxedcell{\texttt{;}} & \boxedcell{\texttt{else}} & \boxedcell{\texttt{\textvisiblespace}} & \boxedcell{\texttt{b}} & \boxedcell{\texttt{=}} & \boxedcell{\texttt{c}} & \boxedcell{\texttt{+}} & \boxedcell{\texttt{d}} & \boxedcell{\texttt{+}} & \boxedcell{\texttt{e}} & \boxedcell{\texttt{;}} \\
\end{tabular}
\end{myboxedtable}

To check if the space character is grammar-insensitive, we test whether inserting additional spaces at the same position affects grammatical validity.  
For example, the following variant — using two consecutive spaces between \boxedcell{\texttt{else}} and \boxedcell{\texttt{b}} is valid:

\begin{myboxedtable}
\begin{tabular}{|*{16}{c|}}
\hline
\boxedcell{if} & 
\boxedcell{(} & 
\boxedcell{a} & 
\boxedcell{)} & 
\boxedcell{;} & 
\boxedcell{else} & 
\boxedcell{\textvisiblespace} & 
\boxedcell{\textvisiblespace} & 
\boxedcell{b} & 
\boxedcell{=} & 
\boxedcell{c} & 
\boxedcell{+} & 
\boxedcell{d} & 
\boxedcell{+} & 
\boxedcell{e} & 
\boxedcell{;} \\
\hline
\end{tabular}
\end{myboxedtable}~\cmark{}

This indicates that the space character is grammar-insensitive and can be used to confirm or refine token boundaries.
Next, we test inserting a space at each candidate boundary. Boundaries where a space can be inserted are considered valid boundaries; if a space cannot be inserted, the adjacent two segments need to be merged.

\subsubsection{Token Classification}

\textbf{Token classification aims to identify which strings belong to the same token class.} For example, in TinyC, \emph{a} and \emph{b} are both identifiers and thus belong to the same class. The purpose of this classification is to reduce the number of distinct token types, thereby simplifying the grammar inference by reducing the variety of tokens that need to be processed.
Once the input is segmented into raw tokens, the next step is to group tokens that belong to the same class. This is achieved using a decision tree classifier, which leverages the distributional behavior of tokens in different contexts.

Suppose the following examples have already been decomposed and used to construct an initial decision tree.

\begin{myboxedtable}
\begin{tabular}{|*{1}{c|}}
\hline
\boxedcell{;}  \\
\hline
\end{tabular}
\end{myboxedtable}

\begin{myboxedtable}
\begin{tabular}{|*{2}{c|}}
\hline
\boxedcell{c}\boxedcell{;}  \\
\hline
\end{tabular}
\end{myboxedtable}

\begin{myboxedtable}
\begin{tabular}{|*{4}{c|}}
\hline
\boxedcell{b}\boxedcell{=}\boxedcell{c}\boxedcell{;}  \\
\hline
\end{tabular}
\end{myboxedtable}

Next, we decompose a new example:
\begin{myboxedtable}
\begin{tabular}{|*{4}{c|}}
\hline
\boxedcell{b}\boxedcell{+}\boxedcell{c}\boxedcell{;}  \\
\hline
\end{tabular}
\end{myboxedtable}

We use the new example for lexical inference. Specifically, we iterate over each token, insert it into the root of the decision tree, and adjust the tree to satisfy the following properties:
% Properties required for the decision tree:
For each leaf node, all strings must exhibit consistent behavior under the given context set \textit{C}.
For each internal node, strings in the left child must not be accepted in the current node’s context, while strings in the right child must be accepted.
Here, the context set \textit{C} refers to all the contexts derived from the currently selected short examples used for lexical inference (not the contexts from the original examples).
The context of a token is defined as a pair consisting of all tokens that precede it and all tokens that follow it. For example, the context of token c is $\square;$ and $b=\square;$, which we represent as: Con(c) = $\{ \square;,\, b=\square; \}$.

\begin{figure}
\setlength{\abovecaptionskip}{2pt} 
\setlength{\belowcaptionskip}{2pt} 
    \centering
    \includegraphics[width=0.6\linewidth]{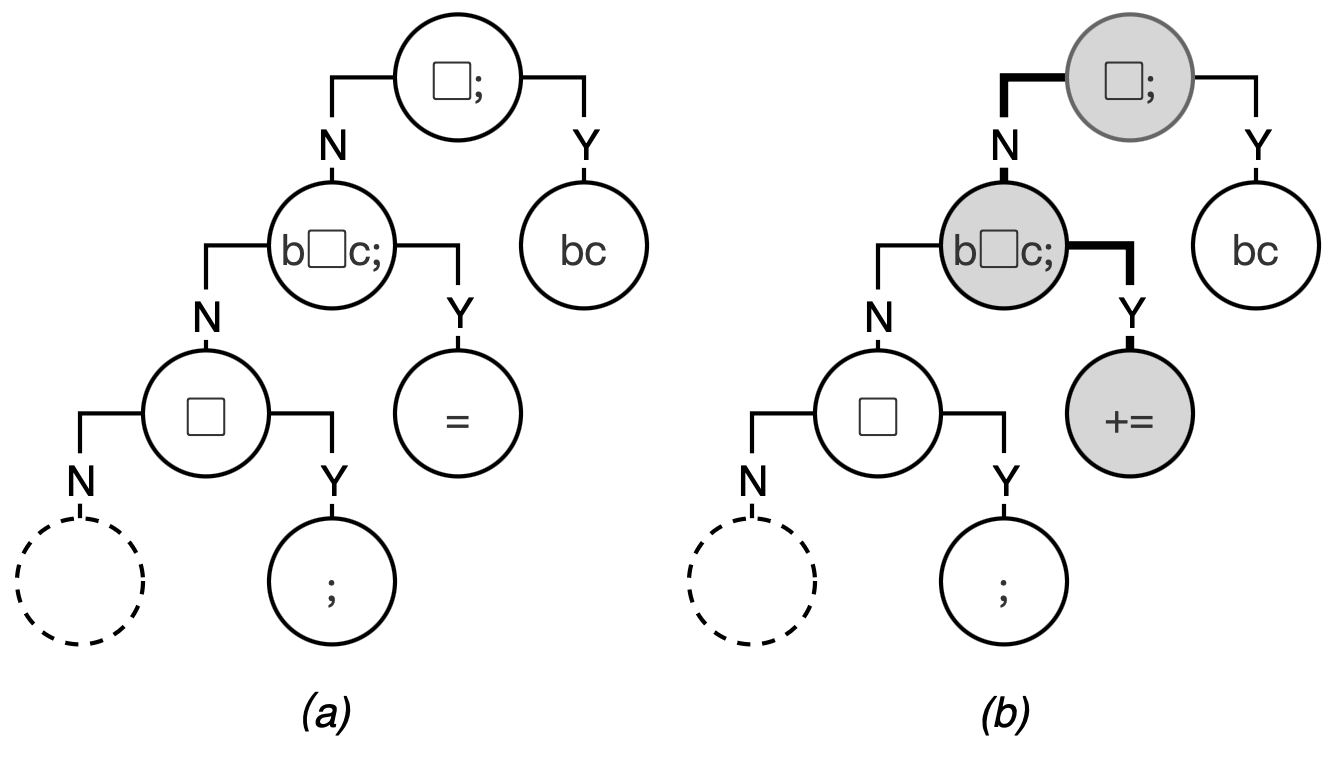}
    \caption{Decision Tree}
    \label{fig:decision_tree}
\end{figure}

\begin{algorithm}[t]
\footnotesize
\caption{Decision Tree Update Procedures}\label{alg:classification}
\begin{algorithmic}[1]
\STATE \textbf{Input:} decision tree $T$,examples $E$, token value $V$
\STATE \textbf{Output:} updated decision tree
\Function{ClassifyToken}{T,E,V}
    \STATE $T.v\leftarrow T.v \cup \{V\}$ \Comment{Add the value to the root node}
    \STATE $T_1\gets$ \Call{UpdateTree}{T,E}\Comment{Adjust the decision tree}
    \STATE \Return{$T_1$}
\EndFunction
\vspace{1ex}
\Function{UpdateNode}{T,E}
    \IF{T is leaf}
        \STATE $C\leftarrow$\Call{Con}{E}
        \STATE c $\gets$ \Call{CheckContexts}{T.v,C}\Comment{Check whether all contexts behave consistently}
        \IF{c=None}
            \RETURN
        \ENDIF
    \ELSE
        \STATE $c\leftarrow T.c$
    \ENDIF
    \STATE $T.r.v\leftarrow \{i\in T.v|O(i\odot c)\}$\Comment{Split the node using context.}
    \STATE $T.l.v\leftarrow V\setminus T.r.v$
    \STATE \Call{UpdateNode}{T.l}
    \STATE \Call{UpdateNode}{T.r}
\EndFunction
\end{algorithmic}
\end{algorithm}

We present the pseudocode for this step in Algorithm \ref{alg:classification}. The key assumption is that tokens within the same class have identical contextual distributions, meaning they can be used interchangeably in the same grammatical contexts. Using this principle, we build a decision tree where internal nodes represent context-based distinctions, and leaves correspond to token classes.

\subsubsection{Character-Level Generalization}
\textbf{Character-level generalization enhances the inferred grammar’s generality.}
Having clustered tokens, we proceed to generalize the character-level representation of each token class. Since tokens are typically described by regular languages, we adopt the L* algorithm~\cite{angluin1987learning} to infer a deterministic finite automata (DFA) for each token class. 

Consider the following example:

\begin{myboxedtable}
\begin{tabular}{|*{1}{c|}}
\hline
\boxedcell{;}  \\
\hline
\end{tabular}
\end{myboxedtable}

\begin{myboxedtable}
\begin{tabular}{|*{2}{c|}}
\hline
\boxedcell{c}\boxedcell{;}  \\
\hline
\end{tabular}
\end{myboxedtable}

\begin{myboxedtable}
\begin{tabular}{|*{4}{c|}}
\hline
\boxedcell{b}\boxedcell{=}\boxedcell{c}\boxedcell{;}  \\
\hline
\end{tabular}
\end{myboxedtable}

Our decision tree is shown in Figure 2(a), where b and c are grouped into the same class , and = and ; are each classified separately. The decision tree only performs classification and cannot generalize to unseen data, so it cannot fully represent lexical rules and therefore cannot serve as an oracle for equivalence queries. We need to convert each leaf node into an automaton to recognize the strings of that token. Each leaf node contains a set of strings S, whose string space is determined by all contexts C from the root to the leaf: a string belongs to the token if its behavior in C matches the strings in S.

To generalize {b, c}, we use the L* algorithm. The original oracle F(x) only checks the correctness of a single string. We wrap it into a token oracle G(x, C) that determines whether x behaves consistently with S in the context C, thus checking if x belongs to that token. Based on this, we construct an approximate equivalence query oracle H(X): it samples from a hypothesis X, returns strings rejected by G (estimating X-T), and checks whether all strings in S are included in X (estimating T-X). While full equivalence is not guaranteed, experience shows this is sufficient to infer complete lexical rules.

When generalizing with L*, a charset must be provided: if the token contains a letter, all letters are included; if it contains digits, all digits are included. This allows {b, c} to be generalized to the entire identifier space.

Finally, by leveraging the DFA inferred for each token, we can construct a complete set of lexical rules. These lexical rules, when combined with the inferred grammar rules, form a full parser capable of analyzing input strings.
\subsubsection{Lexical Validation}
\textbf{Since lexical rules are inferred from short examples, they may fail on complex inputs, thus requiring validation.} To verify their applicability, we first use the current lexical rules to parse the original examples. Each token in the original examples is then replaced with its corresponding equivalent, as determined by the lexical inference. The resulting modified examples are validated against the oracle. If an error is detected, we identify the shortest failing example using the \textit{decomposition forest}, incorporate it as a counterexample to refine the lexical rules.

In the example shown in Figure~\ref{fig:decision_tree}, we treat $+$ and $=$ as belonging to the same token class. To verify this rule, we test it on the original examples by substituting $+$ with $=$ and vice versa.

\begin{myboxedtable}
\begin{tabular}{*{15}{c}} % 15列，按你元素数改
\boxedcell{\texttt{if}} & \boxedcell{\texttt{(}} & \boxedcell{\texttt{a}} & \boxedcell{\texttt{)}} & \boxedcell{\texttt{;}} & 
\boxedcell{\texttt{else}} & \boxedcell{\texttt{b}} & \boxedcell{\texttt{=}} & \boxedcell{\texttt{c}} & \boxedcell{\texttt{+}} & 
\boxedcell{\texttt{d}} & \boxedcell{\texttt{+}} & \boxedcell{\texttt{e}} & \boxedcell{\texttt{;}} \\
\end{tabular}
\end{myboxedtable}~\cmark{}

\begin{myboxedtable}
\begin{tabular}{*{15}{c}} % 15列，按你元素数改
\boxedcell{\texttt{if}} & \boxedcell{\texttt{(}} & \boxedcell{\texttt{a}} & \boxedcell{\texttt{)}} & \boxedcell{\texttt{;}} & 
\boxedcell{\texttt{else}} & \boxedcell{\texttt{b}} & \boxedcell{\texttt{=}} & \boxedcell{\texttt{c}} & \boxedcell{\texttt{+}} & 
\boxedcell{\texttt{d}} & \boxedcell{\texttt{=}} & \boxedcell{\texttt{e}} & \boxedcell{\texttt{;}} \\
\end{tabular}
\end{myboxedtable}~\xmark{}

As one can see, the modified inputs are rejected by the oracle, indicating that the presumed equivalence is invalid. We then traverse the \textit{decomposition forest} to locate a shorter counterexample exhibiting this failure and incorporate them into lexical inference, as shown in the following example.

\begin{myboxedtable}
\begin{tabular}{|*{6}{c|}}
\hline
\boxedcell{c}\boxedcell{+}\boxedcell{d}\boxedcell{+}\boxedcell{e}\boxedcell{;}  \\
\hline
\end{tabular}
\end{myboxedtable}

\begin{algorithm}[t]
\footnotesize
\caption{Distributional Matrix Based Grammar Inference}\label{alg:grammar_infer}
\begin{algorithmic}[1]
\STATE \textbf{Input:} examples $E$
\STATE \textbf{Output:} context free grammar
\Function{InferGrammar}{$E$}
    \STATE $M \leftarrow$ \Call{BuildDistributionalMatrix}{$E$}
    \STATE $T \leftarrow$ \Call{BuildFlatParseTrees}{$E$}
    \WHILE{true}
        \STATE $G \leftarrow$ \Call{BuildSwapGraph}{$M,T$}
        \STATE $C \leftarrow$ \Call{GetMaximalCliques}{$G$}
        \STATE $C_{merge} \leftarrow$ \Call{SelectBestClique}{$C, C_e$} 
        \IF{$C_{merge}=None$} 
            \STATE break \Comment{If no generalizable maximal clique is found,terminate the inference.}
        \ENDIF
        
        \STATE \Call{mergeBubbles}{$T,C_{merge}$} \Comment{Repeatedly merge bubbles to generalize the grammar.} 
    \ENDWHILE
    \STATE \Return \Call{ToGrammar}{$T$}
\EndFunction
\end{algorithmic}
\end{algorithm}

\subsection{Distributional Matrix Based Grammar Inference}

In this section, we detail our methodology for grammar inference based on distributional matrix. The process begins with an empty grammar, which is incrementally constructed by identifying and generalizing short examples from the \textit{decomposition forest}. This iterative process continues, progressively incorporating new short examples to generalize and expand the grammar until all training examples are covered.

Algorithm~\ref{alg:grammar_infer} outlines the steps of our method. We first construct a distributional matrix from the given examples. The distributional matrix serves as an observation table~\cite{clark2016distributional,adriaans1999learning,angluin1987learning}. Then, we build a flat parse tree for each sample, i.e., a parse tree whose root node is the start symbol, and whose child nodes are all the tokens of the sample. Based on these trees, we construct a swap graph, where each node represents a \textit{bubble}, and edges indicate that two \textit{bubbles} can be swapped. Finally, we identify maximal cliques in the graph, select one that contains generalized examples, and merge the corresponding bubbles. The process is then repeated.

This method guarantees first-order swap correctness. Notably, the oracle is queried only once - during the initial construction of the distributional matrix. All subsequent steps are oracle-free.

To formally describe the proposed \textit{Distributional Matrix Based Grammar Inference}, we introduce the following notation.
A \emph{substring} refers to any contiguous fragment $s_i$ extracted from the input strings.  For example, the substrings of $abc$ are $a$, $b$, $c$, $ab$, $bc$, and $abc$.
A \emph{context} is denoted as $p \, \square \, q$, where $p$ and $q$ are strings that represent the prefix and suffix surrounding a missing fragment.   For example, the contexts of $abc$ are $\square$, $\square c$, $a\square$, $\square bc$, $a\square c$ and $ab\square $.
We use the notation $a \odot (b \square c)$ to denote the concatenation of strings $a$, $b$, and $c$, \textit{i.e.}, $a \odot b \square c = abc$.
For a context set $C$ and a substring set $S$, we define  $C \odot S = \{\, l u r \mid l \square r \in C,\, u \in S \,\}$.

\textbf{First-order Swap Correctness}: Let  $T = \{t_1, t_2, \ldots, t_n\}$ be a set of parse trees.  
For each parse tree $t_i$, let $NT(t_i)$ denote the set of nonterminal symbols it contains.  
The complete set of nonterminal symbols across all parse trees is given by $N = \bigcup_{t_i \in T} NT(t_i)$.
For a nonterminal symbol $x$, let $\mathsf{Con}(x)$ denote the set of all contexts in which $x$ appears across the parse trees, and let $\mathsf{Sub}(x)$ denote the set of all subsequences derived from $x$. 
By combining the contexts and subsequences, we obtain a set of examples  
$S = \{ \mathsf{Sub}(x) \odot \mathsf{Con}(x) \mid x \in N \}$,  
where $\odot$ denotes a concatenation operation.
We say that a set of parse trees satisfies \textit{first-order swap correctness} if all examples in the set $S$ are valid.

\subsubsection{Distributional matrix construction}
\textbf{The distributional matrix encodes distributional information from the oracle, allowing grammar inference without requiring additional queries.}
Each row of the distributional matrix $M$ corresponds to a substring $s_i$, and each column corresponds to a context $c_i$.

For each substring $s_i$ and context $c_j$, we define $M[i, j] = 1$ if and only if inserting $s_i$ into the hole yields a valid string:
{
\vspace{-0.92em}
\[
M[i, j] =
\begin{cases}
1 & \text{if } \text{oracle}(s_i\odot c_j) = \texttt{accept} \\
0 & \text{otherwise}
\end{cases}
\]
\vspace{-0.92em}
}

To construct $M$, we enumerate all substrings $s_i$ in the input examples and all possible contexts $p \, \square \, q$ by deleting a single contiguous fragment from each example string.

\textbf{Example:} Suppose we have already decomposed the following examples, which are to be used for grammar inference.

\begin{myboxedtable}
\begin{tabular}{|*{1}{c|}}
\hline
\boxedcell{;}  \\
\hline
\end{tabular}
\end{myboxedtable}

\begin{myboxedtable}
\begin{tabular}{|*{2}{c|}}
\hline
\boxedcell{a}\boxedcell{;}  \\
\hline
\end{tabular}
\end{myboxedtable}

\begin{myboxedtable}
\begin{tabular}{|*{4}{c|}}
\hline
\boxedcell{(}\boxedcell{a}\boxedcell{)}\boxedcell{;}  \\
\hline
\end{tabular}
\end{myboxedtable}

The distributional matrix $M$ corresponding to these substrings and contexts is shown below. All substrings, along with the contexts obtained by deleting contiguous substrings from the input examples, serve as the rows and columns of the distributional matrix.
{
\footnotesize
\[
\begin{array}{c|cccccc}
\text{Sub} \backslash \text{Con} & \square & \square ; & (\square); & a\square & (a) \square \\ \hline
; & 1 & 0 & 0 & 1 & 1 \\
a & 0 & 1 & 1 & 0 & 0 \\
a; & 1 & 0 & 0 & 0 & 0 \\
(a) & 0 & 1 & 1 & 0 & 0 \\
(a); & 1 & 0 & 0 & 0 & 0 \\
\end{array}
\]
}

\subsubsection{Flat parse tree construction.}
\textbf{For each input sample, we construct a flat parse tree that serves as the initial structure for grammar inference.} A flat parse tree contains no hierarchical structure, each token in the sample is treated as an atomic unit and placed directly under the root.

\begin{figure}
    \centering
    \includegraphics[width=0.7\linewidth]{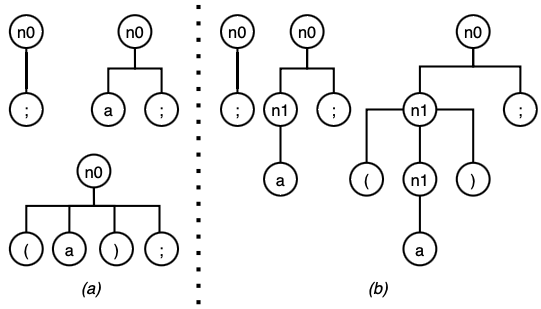}
    \caption{(a) flat parse trees~(b) generalized parse trees}
    \label{fig:syntax-trees}
\end{figure}

The structures of the flat parse trees are shown in Figure~\ref{fig:syntax-trees}(a).
These trees provide a structural foundation for later generalization and abstraction. Subsequently, we will derive new grammar structures by merging bubbles within these trees.
\subsubsection{Construct swap graph}

\textbf{The swap graph identifies swappable bubbles for merging, enabling grammar generalization.}
The swap graph is an undirected graph where each node represents a bubble in the parse tree, and each edge indicates whether two bubbles can be swapped. Each bubble corresponds to a contiguous subtree that may potentially be abstracted into a nonterminal. Through the swap graph, we can identify which bubbles can be merged without violating first-order swap correctness. We construct the swap graph solely based on the distributional matrix, without any additional queries to the oracle.

\begin{figure}
    \centering
    \includegraphics[width=0.8\linewidth]{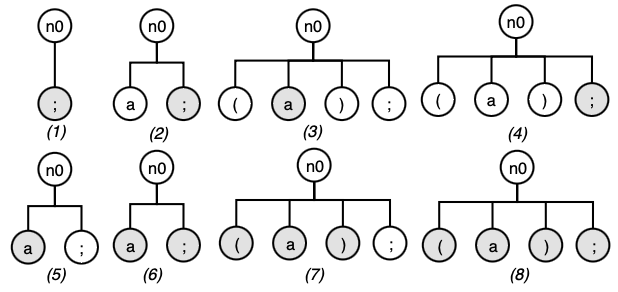}
    \caption{Bubbles. Each bubble corresponds to several shadowed nodes in the tree.}
    \label{fig:bubbles}
\end{figure}

Each bubble can be mapped to a set of contexts and a set of substrings. More specifically:
\begin{itemize}[leftmargin=*]
    \item If the bubble is a nonterminal, its substrings are defined as the union of the strings of all bubbles associated with that nonterminal. Otherwise, its substring is simply the string at the corresponding position in the sample.
    \item If the bubble’s parent is a nonterminal, its context is defined as the union of the contexts of all bubbles associated with that nonterminal parent. Otherwise, its context is simply the context observed at its position in the original sample.
\end{itemize}
The contexts and substrings corresponding to each bubble in Figure~\ref{fig:bubbles} are listed below.
Since a flat parse tree is used in this case, each bubble corresponds to exactly one substring and one context.
{\footnotesize
\[
\begin{array}{c|cc}
\text{Bubble} & \text{Sub} & \text{Con} \\ \hline
1 & ; & \square \\
2 & ; & a\square \\
3 & a & (\square); \\
4 & ; & (a)\square \\
\end{array}
\quad
\begin{array}{c|cc}
\text{Bubble} & \text{Sub} & \text{Con} \\ \hline
5 & a & \square; \\
6 & a; & \square \\
7 & (a) & \square; \\
8 & (a); & \square \\
\end{array}
\]
}

\begin{figure}
    \centering
    \includegraphics[width=0.5\linewidth]{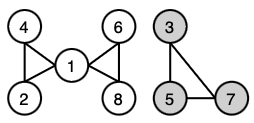}
    \caption{Swap Graph. Shadowed nodes represent the maximal clique we selected.}
    \label{fig:swap_graph}
\end{figure}

By querying the distributional matrix, we check whether all substrings of two bubbles can be placed into all of their associated contexts. If so, we add an undirected edge between them in the swap graph. The constructed swap graph is shown in Figure~\ref{fig:swap_graph}.

When a group of nodes forms a clique, \textit{i.e.}, every pair of nodes is connected, it can be safely merged into a shared nonterminal without violating first-order swap correctness.

\subsubsection{Bubble merging}
\textbf{To generalize the grammar, we aim to merge swappable \emph{bubbles} into shared nonterminals.}
The process begins by identifying \emph{maximal cliques} in the swap graph. Each maximal clique represents a set of bubbles in which every pair is connected, indicating that they satisfy first-order swap correctness.

We enumerate all maximal cliques in descending order of size.
Once we find a maximal clique whose combination of contexts and substrings produces a sample that generalizes the current grammar (\textit{i.e.}, a sample not already covered by the grammar), as shown in Figure~\ref{fig:swap_graph}, we merge the clique.
The result of the merge is showed in Figure~\ref{fig:syntax-trees}(b).
After each merge, the swap graph is reconstructed, and the process continues with further bubble selections.
If no valid maximal clique is found, the generalization process terminates.
However, this process only ensures \emph{first-order swap correctness}; the resulting grammar may not achieve perfect precision.
\section{Evaluation}

The ultimate goal of our proposed approach is to automatically infer formal grammars from a set of given programs, particularly for programming languages with numerous and intricate grammar rules. To evaluate whether this goal has been achieved, we applied \tool{} to eight micro-benchmarks, three macro-benchmarks, and five new benchmarks consisting of programs written in complex programming languages, and compared its performance against state-of-the-art approaches. We then conducted an in-depth analysis of the results to answer the following three research questions.
    
\textbf{RQ1} Grammar Quality: How accurate are the grammars inferred by our method, in terms of their ability to accept valid inputs and reject invalid ones?
    
\textbf{RQ2} Efficiency: How efficient is our method compared to prior tools in terms of runtime on large and complex datasets?
    
\textbf{RQ3} Readability: Are the inferred grammars concise?

\subsection{Experimental Setup}
To evaluate the effectiveness and efficiency of \tool{} on complex grammars, we applied \tool{} to a constructed benchmark dataset containing 16 grammars, including both simple and complex grammars from high-level programming languages. Specifically, we include all the original datasets from \sotak
\footnote{\url{https://github.com/Sinpersrect/kedavra}}, which represents the state-of-the-art in grammar inference. As the grammars in \sotak{}'s benchmark are relatively simple, we augmented the dataset with five additional grammars corresponding to C, Lua, Java, Rust, and C++, selected to reflect the complexity of real-world programming languages. The training and test programs for these five grammars were sourced from a real-world dataset, \emph{The Algorithms}\footnote{\url{https://github.com/thealgorithms}}. Oracles for these grammars were constructed using grammars from the \emph{ANTLR4 grammars-v4}\footnote{\url{https://github.com/antlr/grammars-v4}}  repository. Since comments in programming languages are generally easy to handle and follow standardized formats, and considering that comments typically do not affect the grammatical correctness of code, we remove all comments from the dataset.

Hence, the overall dataset for the experiments contains 31 training datasets across 16 grammars. Table~\ref{table:dataset-complexity} summarizes the grammar sizes, with complex grammars averaging 165 non-terminal symbols compared to the \sotak{}'s benchmark only contains 7 non-terminal symbols. The dataset statistics are further summarized in Table~\ref{table:dataset}. For complex grammars (i.e., C, Lua, Java, Rust, and Cpp), the datasets contain an average of 98 examples, with an average length of 2008 characters per example. For simple grammars, the datasets contain an average of 29 examples, with an average length of 63 characters. These differences reflect the variation in grammar complexity and dataset scale, providing a basis for evaluating \tool{} across varying grammar complexities. The benchmark dataset has been made publicly available on GitHub\footnote{\url{https://github.com/Sinpersrect/gi-benchmarks}} for future use.

All experiments were conducted on a Linux server equipped with an Intel(R) Core(TM) i9-14900F CPU and 64GB RAM, with a timeout threshold of 48 hours imposed for each experiment. For each approach, all language experiments (11 simple + 5 complex) were executed simultaneously on the same server, with identical settings for each approach to ensure fairness. 

\begin{table}
\setlength{\abovecaptionskip}{2pt} 
\setlength{\belowcaptionskip}{2pt} 
\footnotesize
\caption{Grammar size of Dataset: NT/T = unique (non-) terminals; A = rules (alternatives); l(A) = avg. rule length; S = sum of rule lengths.}
\label{table:dataset-complexity}
\begin{tabular}{l|lllll}
\toprule
Name & NT & T & A & l(A) & S \\
\midrule
arith & 4 & 8 & 9 & 1.56 & 14 \\
math & 6 & 14 & 16 & 2.06 & 33 \\
fol & 8 & 16 & 18 & 2.89 & 52 \\\hline
json & 7 & 12 & 17 & 1.76 & 30 \\
lisp & 5 & 7 & 11 & 2.27 & 25 \\
turtle & 4 & 10 & 8 & 3.12 & 25 \\
while & 4 & 10 & 8 & 3.12 & 25 \\
xml & 7 & 10 & 19 & 2.58 & 49 \\\hline
curl & - & - & - & - & - \\
tinyc & 9 & 13 & 18 & 2.28 & 41 \\
minic & 17 & 21 & 35 & 2.23 & 78 \\\hline
c & 91 & 113 & 320 & 1.88 & 601 \\
cpp & 228 & 126 & 624 & 1.75 & 1,095 \\
java & 210 & 124 & 534 & 1.72 & 917 \\
lua & 45 & 64 & 135 & 1.90 & 257 \\
rust & 253 & 115 & 706 & 1.71 & 1,205 \\
\bottomrule
\end{tabular}
\end{table}

\begin{table}
\setlength{\abovecaptionskip}{2pt}
\setlength{\belowcaptionskip}{2pt} 
\footnotesize
\caption{Example strings S with their avg char size; \# = nr. programs.}\label{table:dataset}
\begin{tabular}{c|cc|cc|cc|cc|cc}
  \hline
  & \multicolumn{2}{c|}{R0} & \multicolumn{2}{c|}{R1}  & \multicolumn{2}{c|}{R2} & \multicolumn{2}{c|}{R5}  & \multicolumn{2}{c}{Test} 
   \\
    & \# & avg & \# & avg & \# & avg & \# & avg & \# & avg  \\\hline
arith&17&2.3&\multicolumn{6}{c|}{n/a}&1,000&65.9\\
fol&36&14.5&\multicolumn{6}{c|}{n/a}&1,000&106.5\\
math&62&5.5&\multicolumn{6}{c|}{n/a}&1,000&72.1\\\hline
json&71&3.9&30&11.7&30&8.6&\multicolumn{2}{c|}{n/a}&1,000&10.8\\
lisp&26&2.5&30&79.2&30&24.8&\multicolumn{2}{c|}{n/a}&1,000&73.2\\
turtle&33&7.7&35&41.1&35&25.6&\multicolumn{2}{c|}{n/a}&1,000&34.6\\
while&10&15.5&30&171.4&30&217&\multicolumn{2}{c|}{n/a}&1,000&215.2\\
xml&40&11.6&20&27.8&20&27.5&\multicolumn{2}{c|}{n/a}&1,000&25\\\hline
curl&25&20.7&25&22.1&25&22.0&\multicolumn{2}{c|}{n/a}&1,000&22.2\\
tinyc&25&80.5&25&96.2&25&86.4&10&514&1,000&94.2\\
minic&10&107&\multicolumn{2}{c|}{n/a}&\multicolumn{2}{c|}{n/a}&\multicolumn{2}{c|}{n/a}&1,000&267.3\\\hline
c&81&1,175&\multicolumn{6}{c|}{n/a}&326&1,488\\
cpp&70&2,493&\multicolumn{6}{c|}{n/a}&283&2,610\\
java&254&1,730&\multicolumn{6}{c|}{n/a}&1,019&1,738\\
lua&30&1,822&\multicolumn{6}{c|}{n/a}&120&1,276\\
rust&57&2,821&\multicolumn{6}{c|}{n/a}&232&2,776\\\hline
\end{tabular}
\end{table}

\subsection{RQ1: Grammar Quality}
\begin{figure}
\setlength{\abovecaptionskip}{2pt} 
\setlength{\belowcaptionskip}{2pt} 
    \centering
    \includegraphics[width=1\linewidth]{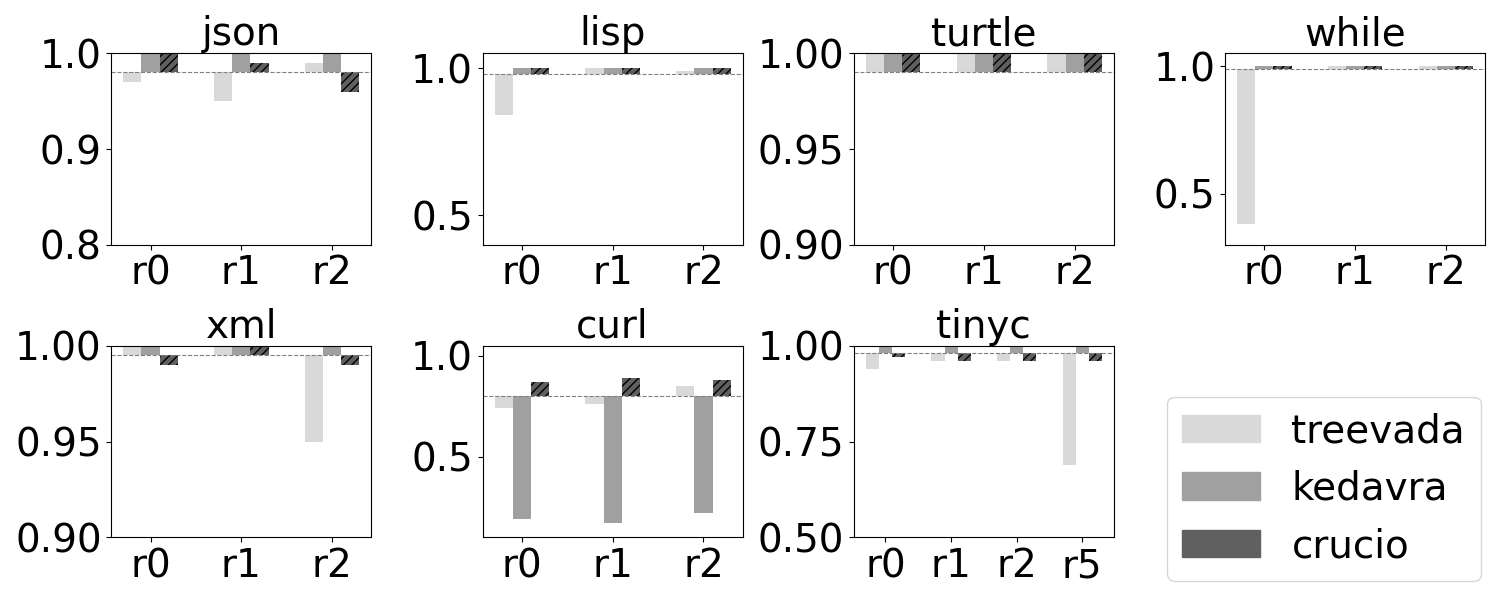}
    \caption{avg F1 score of 10 runs of \sotat{}, \sotak{} and \tool{} on each dataset (R0, R1, R2, R5).}
    \label{fig:stability}
\end{figure}
\begin{table}
\setlength{\abovecaptionskip}{2pt} 
\setlength{\belowcaptionskip}{2pt} 
\scriptsize
\caption{Rerun \sotat{} (left), rerun \sotak{} (middle) and \tool{}(right) on the same example strings; \sotak{}/\tool{} values are average over 10 runs; f1 = F1 score; ± = standard deviation; bold = \tool{} better than or equal to \sotat{}\& \sotak{}.}\label{tab:Grammar_Quality_Comparison}
\begin{tabular}{l|ccc|ccc|ccc}
\hline
  \multicolumn{1}{c|}{}& \multicolumn{3}{c|}{\sotat{}}& \multicolumn{3}{c|}{\sotak{}}  & \multicolumn{3}{c}{\tool{}} \\
   & p & r & f1 & p & r & f1 & p & r & f1 \\
   \hline

arith & 1 & 1 & 1 & 1.0±.0 & 1.0±.0 & 1.0±.0 & \textbf{1.0±.0} & \textbf{1.0±.0} & \textbf{1.0±.0} \\
math & 1 & .47 & .64 & .97±.04 & .76±.2 & .84±.15 & .79±.02 & \textbf{.9±.02} & \textbf{.84±.02} \\
fol & 1 & .53 & .69 & 1.0±.01 & .71±.25 & .81±.16 & \textbf{1.0±.0} & \textbf{1.0±.0} & \textbf{1.0±.0} \\\hline
json & 1 & .94 & .97 & 1.0±.0 & .96±.02 & .98±.01 & \textbf{1.0±.0} & \textbf{1.0±.0} & \textbf{1.0±.0} \\
lisp & .99 & .73 & .84 & 1.0±.0 & 1.0±.0 & 1.0±.0 & \textbf{1.0±.0} & \textbf{1.0±.0} & \textbf{1.0±.0} \\
turtle & 1 & 1 & 1 & 1.0±.0 & 1.0±.0 & 1.0±.0 & \textbf{1.0±.0} & \textbf{1.0±.0} & \textbf{1.0±.0} \\
while & 1 & .24 & .38 & 1.0±.0 & 1.0±.0 & 1.0±.0 & \textbf{1.0±.0} & \textbf{1.0±.0} & \textbf{1.0±.0} \\
xml & 1 & 1 & 1 & 1.0±.0 & 1.0±.0 & 1.0±.0 & \textbf{1.0±.0} & .99±.0 & .99±.0 \\\hline
curl & .95 & .6 & .74 & .91±.12 & .12±.14 & .19±.2 & .85±.01 & \textbf{.88±.01} & \textbf{.87±.01} \\
tinyc & .98 & .9 & .94 & 1.0±.0 & 1.0±.0 & 1.0±.0 & .93±.02 & \textbf{1.0±.0} & .97±.01 \\
minic & .96 & .48 & .64 & .89±.03 & .51±.0 & .65±.01 & \textbf{.98±.0} & \textbf{1.0±.0} & \textbf{.99±.0} \\\hline
c & - & - & - & - & - & - & \textbf{.65±.04} & \textbf{.85±.03} & \textbf{.73±.04} \\
cpp & - & - & - & - & - & - & \textbf{.62±.02} & \textbf{.82±.02} & \textbf{.7±.02} \\
java & - & - & - & - & - & - & \textbf{.75±.04} & \textbf{.93±.01} & \textbf{.83±.03} \\
lua & - & - & - & - & - & - & \textbf{.59±.05} & \textbf{.83±.05} & \textbf{.69±.04} \\
rust & - & - & - & - & - & - & \textbf{.42±.04} & \textbf{.78±.03} & \textbf{.54±.03} \\

\bottomrule
\end{tabular}
\end{table}
Our first research question focuses on assessing the quality of the grammar produced by \tool{} and how it compares to existing tools. Consistent with prior works, we adopt precision, recall, and F1 score to measure grammar quality. At the same time, we introduce a more reliable precision measurement (i.e., \emph{swap precision}) to overcome key limitations of sampling-based precision measurement (i.e., \emph{sampling precision}) from existing approaches, specifically their inability to account for structural bias and sensitivity to sample length distributions.

All existing methods \sotaa{}, \sotat{}, and \sotak{} evaluate precision using sampling-based estimation. However, this approach suffers from two major limitations:
\begin{itemize}[leftmargin=*]
    \item \underline{Structural bias:} \emph{sampling precision} measures the proportion of valid examples generated from a grammar using a sampling algorithm. However, this metric can be structurally manipulated. For example, one can deliberately construct a grammar where most production rules are designed to generate a small set of real examples. As shown in Table~\ref{table:sample-swap}, the sampling process in such cases almost always produces these few examples, leading to a sampling precision of 1. By additionally introducing a few overly general rules (e.g., matching arbitrary strings), recall can also reach 1. In practice, such grammars may deceptively achieve an F1 score of 1 under the evaluation schemes used by \sotaa{}, \sotat{}, and \sotak{}, despite severely overfitting the training data and incorporating many invalid generalizations. Consequently, these grammars do not represent valid or meaningful abstractions of the underlying language. 
    \item \underline{Sample length bias:} During sampling, the length of generated examples is typically uncontrollable and is influenced by the grammar structure itself. Grammars that favor shorter examples tend to produce strings that are more likely to match real inputs, resulting in artificially high sampling precision. 
    This shows that sampling precision is confounded not only by a grammar's expressiveness but also by the length distribution of generated samples, making it unreliable as an objective metric for assessing grammar quality. For complex grammars, it is particularly easy for the sampling process to generate extremely long and incorrect samples, further diminishing the reliability of this metric.
\end{itemize}

To address this, we propose a new precision evaluation method: Swap Precision. 

\textbf{Swap Precision}: Given some examples $E = \{e_1, e_2, \ldots, e_n\}$, we can construct a parse tree for each example, forming the set $T = \{t_1, t_2, \ldots, t_n\}$.  
The set of all nonterminal symbols in the parse tree $t_i$ is denoted by $NT(t_i)$.  
The complete set of nonterminal symbols is denoted by $N = \bigcup_{t_i \in T} NT(t_i)$.

We denote $\mathsf{Con}(x)$ as the set of all contexts in which the nonterminal symbol $x$ appears across all parse trees, and $\mathsf{Sub}(x)$ as the set of all subsequences derived from $x$.  
By combining the contexts and subsequences, we obtain a set of examples  
$S = \{ \mathsf{Sub}(x) \odot \mathsf{Con}(x) \mid x \in N \}$,  
where $\odot$ denotes a composition operation.

The swap precision is defined as:
\begin{equation}
    p_{\mathrm{swap}} = \frac{|\{ s \in S \mid s \in L(G) \}|}{|S|}
\end{equation}

Validating each example in $S$ is time-consuming. Therefore, if $|S| > 1000$, we randomly select 1,000 examples from $S$ for evaluation. To evaluate swap precision, we use the subset of test set that can be parsed by the current grammar.

This provides a more reliable measure of grammar precision by evaluating the correctness of recognizing equivalent substrings over the distribution of the test set. Accordingly, in the following experiments, we use \emph{swap precision} instead of \emph{sampling precision} to better reflect grammar quality. We define all metrics as follows:

\underline{\emph{Swap Precision:}} Here, \emph{swap precision} is defined as the proportion of grammatically valid strings obtained by swapping substrings corresponding to the same non-terminal symbols in parse trees derived from successfully parsed test set.
\underline{\emph{Recall:}}
We compile the inferred grammar into a parser and count how many of the given test programs that parser accepts.
Programs with a parsing time exceeding 60 seconds are considered parsing failures.
\underline{\emph{F1 score:}}
The F1 score is the harmonic mean of precision and recall.

\begin{table}
\setlength{\abovecaptionskip}{2pt} 
\setlength{\belowcaptionskip}{2pt} 
\footnotesize
\caption{Sampling precision and swap precision performance on the invalid grammars we constructed.}
\label{table:sample-swap}
\begin{tabular}{l|lllll}
\toprule
Name & $p_{sample}$ & $p_{swap}$ & r & $f1_{sample}$ & $f1_{swap}$ \\
\midrule
arith & 1 & 0.089 & 1 & 1 & 0.16 \\
math & 1 & 0.063 & 1 & 1 & 0.12 \\
fol & 1 & 0.024 & 1 & 1 & 0.05 \\\hline
json & 1 & 0.035 & 1 & 1 & 0.07 \\
lisp & 1 & 0.018 & 1 & 1 & 0.04 \\
turtle & 1 & 0.028 & 1 & 1 & 0.05 \\
while & 1 & 0.009 & 0.987 & 0.99 & 0.02 \\
xml & 1 & 0.013 & 1 & 1 & 0.03 \\\hline
curl & 1 & 0.894 & 1 & 1 & 0.94 \\
tinyc & 1 & 0.024 & 1 & 1 & 0.05 \\
minic & 1 & 0.02 & 0.972 & 0.99 & 0.04 \\\hline
c & 1 & 0.035 & 0.89 & 0.94 & 0.07 \\
cpp & 1 & 0.013 & 0.93 & 0.96 & 0.03 \\
java & 1 & 0.05 & 0.98 & 0.99 & 0.1 \\
lua & 1 & 0 & 0.96 & 0.98 & 0 \\
rust & 1 & 0.035 & 0.8 & 0.89 & 0.07 \\
\bottomrule
\end{tabular}
\end{table}

\textbf{Results.}Our experimental results on the R0 dataset are shown in Table~\ref{tab:Grammar_Quality_Comparison}, which presents a comparison between \tool{} and two other SOTA tools. Here, we selected \sotat{}\cite{arefin2024fast} and \sotak{}\cite{li2024incremental} as baselines because they are considered among the most advanced and publicly available grammar inference tools in the community.

Overall, as shown in Table~\ref{tab:Grammar_Quality_Comparison}, \tool{} achieves F1 scores equal to or higher than those of \sotat{} and \sotak{} across 14 grammar datasets (i.e., arith, math, fol, json, lisp, turtle, while, curl, minic, c, cpp, java, lua, and rust), indicating that our tool performs better on most datasets.
For example, \tool{} achieved perfect scores (precision, recall, and F1 all 1.0 ± 0.0) on arith, fol, json, lisp, turtle, and while. Even for complex program input datasets such as c, cpp, java, lua, and rust, \tool{} still maintained strong performance with an average F1 score of 0.698, whereas \sotat{} and \sotak{} failed to infer grammars on these datasets.

Specifically, \sotat{} was terminated due to insufficient memory after running for 103 h, 206 h, 56 h, 126 h, and 337 h on C, C++, Java, Lua, and Rust, respectively. In contrast, \sotak{} was terminated after running for 70 h and 66 h on Java and Rust because the Ubuntu system killed the processes with the largest memory consumption once peak memory usage was exhausted., and did not complete after more than 360 h on C, C++, and Lua. These results highlight the effectiveness of \tool{} in generating high-quality grammars.

Exceptions include the xml and tinyc datasets: on xml, \tool{} achieved an F1 score of 0.99, slightly lower than that of 1.0 by both \sotat{} and \sotak{}; on tinyc, it also underperformed compared to \sotak{}'s perfect score of 1.0.
The lower recall on xml (0.99) is due to the use of the L* algorithm for lexical inference, which infers automata under contextual constraints, resulting in a stricter but less generalizable model. For tinyc, the grammar was constructed to ensure first-order swap correctness rather than being validated through sampling, leading to lower precision compared to \sotat{} and \sotak{}. A detailed discussion of precision differences is provided in Section 5. Despite these few exceptions, \tool{} consistently outperforms state-of-the-art tools, demonstrating its effectiveness in producing high-quality grammars.

Beyond evaluating grammar quality, it is also important to assess whether \tool{} can maintain consistent performance across different datasets. To this end, we further analyzed the F1 scores of the same grammars across multiple datasets (R0, R1, R2, and R5), as shown in Figure~\ref{fig:stability}. This figure presents the F1 scores of the three tools across seven grammar datasets. The deviation of the bars from the horizontal reference line reflects fluctuations in F1 scores. Note that each subplot contains a manually added horizontal line to aid visualization of how the F1 score varies across different datasets for each inference algorithm.

\tool{} exhibits smaller performance variations (i.e., shorter distances from the bars to the horizontal axis), while \sotat{} is more sensitive to different dataset characteristics.
For grammars such as Lisp, Turtle, and while, \tool{} consistently achieves an F1 score of 1.0 across all datasets.
In contrast, \sotat{} shows larger F1 score fluctuations on datasets such as json, lisp, while, xml, and tinyc.
This indicates that \tool{} achieves performance stability on par with \sotak{} across different datasets derived from the same underlying grammar. 
Upon closely comparing the designs of \tool{} and the other two methods, we attribute \tool{}'s greater F1 score stability primarily to its design: like \sotak{}, it performs inference on decomposed examples, but its use of decomposition forests allows for much stronger generalization than prior approaches.
\begin{framed}
\textbf{Answer to RQ1:}
\tool{} demonstrates clear advantages on simple datasets, achieving recall improvements of 1.37$\times$ and 1.19$\times$, and F1 score improvements of 1.21$\times$ and 1.13$\times$ over \sotat{} and \sotak{}, respectively, while incurring only a slight decrease in precision (0.97$\times$ and 0.98$\times$). More notably, \tool{} is uniquely capable of handling complex real-world languages such as C, C++, Java, and Rust, successfully inferring grammars where other SOTA methods either fail to terminate or cannot operate at all.
\end{framed}

\subsection{RQ2: Efficiency}

While we have demonstrated the effectiveness of the inferred grammar, efficiency remains a critical factor, reflecting both the speed and resource demands of the inference process. In this RQ, we evaluate \tool{}'s efficiency by measuring its computational resource consumption and processing time. For a fair comparison, we report metrics consistent with those used in SOTA studies:
\textbf{Runtime}: The time cost in executing the tool. 
\textbf{Oracle Time}: We measure the time cost during the oracle checking process.
\textbf{Oracle Calls}: The number of oracle calls.
\textbf{Peak Memory}: We use the Linux `time' command to measure peak memory usage during grammar inference.

\begin{table*}
\setlength{\abovecaptionskip}{2pt} 
\setlength{\belowcaptionskip}{2pt} 
\footnotesize

\caption{Average \sotat{} \& \sotak{} results over 10 runs on R0; $t$ = runtime; $t_{\text{O}}$ = oracle time; $q$ = queries sent to oracle; $m$ = peak memory usage; $\pm$ = standard deviation; \textbf{bold} = \tool{} better than \sotat{} \& \sotak{}.}\label{tab:computational_resources}
\begin{tabular}{c|cccc|cccc|cccc}
  \hline
  \multicolumn{1}{c|}{}& \multicolumn{4}{c|}{\sotat{}}  & \multicolumn{4}{c|}{\sotak{}} &\multicolumn{4}{c}{\tool{}} 
   \\
    & $t[ks]$ & $t_{O}[ks]$  & $q[k]$ & $m[GB]$ & $t[ks]$ & $t_{O}[ks]$  & $q[k]$ & $m[GB]$& $t[ks]$ & $t_{O}[ks]$  & $q[k]$ & $m[GB]$ \\\hline
arith & .00±.0 & .00±.0 & 0.37 & 0.02 & .00±.0 & .00±.00 & .58±.02 & .08±.01 & \textbf{.00±.0} & \textbf{.00±.0} & 1.43±1.3 & .07±.0 \\
math & .01±.0 & .01±.0 & 6.33 & 0.02 & .84±2.1 & .06±.03 & 17.62±8.91 & .08±.04 & .10±.0 & .09±.0 & 35.66±10.2 & .09±.0 \\
fol & .04±.0 & .03±.0 & 16.25 & 0.03 & .28±.2 & .20±.01 & 57.00±4.20 & .11±.10 & .49±.0 & .47±.0 & 181.67±5.3 & .15±.0 \\\hline
json & .01±.0 & .00±.0 & 3.26 & 0.02 & .06±.1 & .01±.00 & 4.00±1.13 & .11±.05 & .05±.0 & .05±.0 & 20.39±11.1 & .08±.0 \\
lisp & .00±.0 & .00±.0 & 1.84 & 0.02 & .01±.0 & .01±.00 & 2.16±.03 & .12±.03 & .01±.0 & .01±.0 & 2.13±.2 & .08±.0 \\
turtle & .01±.0 & .01±.0 & 5.11 & 0.02 & .02±.0 & .02±.00 & 4.53±.19 & .03±.00 & \textbf{.01±.0} & \textbf{.01±.0} & \textbf{4.31±.1} & .08±.0 \\
while & .01±.0 & .01±.0 & 3.23 & 0.02 & .18±.0 & .14±.03 & 42.96±8.53 & .10±.05 & .12±.0 & .12±.0 & 46.74±.0 & .09±.0 \\
xml & .05±.0 & .02±.0 & 9.48 & 0.03 & .23±.0 & .20±.03 & 62.27±8.28 & .09±.00 & .14±.0 & .13±.0 & 56.06±.7 & .09±.0 \\\hline
curl & .09±.0 & .07±.0 & 23.39 & 0.04 & .11±.0 & .09±.02 & 22.27±3.83 & .03±.00 & .22±.0 & .20±.0 & 45.79±3.4 & .09±.0 \\
tinyc & .65±.0 & .05±.0 & 78.02 & 0.17 & .06±.0 & .03±.00 & 19.83±.46 & .03±.00 & \textbf{.02±.0} & \textbf{.01±.0} & \textbf{8.30±.3} & .08±.0 \\
minic & .95±.0 & .42±.0 & 77.52 & 0.12 & .26±.1 & .13±.02 & 32.91±3.92 & .10±.11 & \textbf{.15±.0} & \textbf{.13±.0} & 44.91±.6 & \textbf{.10±.0} \\\hline
c & - & - & - & - & - & - & - & - & \textbf{10.04±1.4} & \textbf{7.42±1.5} & \textbf{988.45±227.5} & \textbf{.90±.2} \\
cpp & - & - & - & - & - & - & - & - & \textbf{84.94±11.6} & \textbf{49.14±9.6} & \textbf{1349.21±505.5} & \textbf{4.46±.4} \\
java & - & - & - & - & - & - & - & - & \textbf{65.61±15.9} & \textbf{16±.7} & \textbf{1778.05±132.5} & \textbf{2.34±.1} \\
lua & - & - & - & - & - & - & - & - & \textbf{5.14±.8} & \textbf{1.86±.7} & \textbf{402.23±206.1} & \textbf{3.47±.5} \\
rust & - & - & - & - & - & - & - & - & \textbf{78.86±5.9} & \textbf{40.57±1.4} & \textbf{1432.73±108.5} & \textbf{3.30±.7} \\

\hline
\end{tabular}
\end{table*}
Table \ref{tab:computational_resources} presents the resource usage of \sotat{}, \sotak{}, and \tool{} on the R0 dataset. Overall, \tool{} requires an average of 0.12 kiloseconds per program for grammar inference, which is faster than \sotat{} (0.17 kiloseconds per program) and \sotak{} (0.19 kiloseconds per program).
Improving grammar quality often reduces efficiency, as shown by Arefin~\cite{arefin2024fast}. This explains why \tool{} takes longer on the math, fol, json, lisp, while, xml, and curl datasets: \tool{} has stronger generalization capabilities, decomposes more examples for grammar inference, computes distributional matrix, and employs the L* algorithm for lexical inference, which is more precise than previous methods and requires more oracle queries.
However, we argue that such trade-offs between effectiveness and efficiency are inevitable. Moreover, it is worth noting that on the math, fol, json, lisp, while, xml, and curl datasets, the time difference between \tool{} and \sotat{}/\sotak{} is relatively small (only a few hundred seconds), which is entirely acceptable.

When handling more complex datasets such as tinyc, minic, c, cpp, java, lua, and rust, our method still maintains competitive speed. This is primarily due to the use of decomposition forest, which avoid redundant decomposition of examples that already cover the full grammar, and the application of lexical inference on the decomposed examples. Furthermore, \tool{} does not rely on sampling to ensure grammar correctness; instead, it strikes a good balance between generalization and overgeneralization. This indicates that, even if the runtime is slightly longer for certain grammars, \tool{} remains practical and usable due to its superior grammar quality and relatively low overall time cost.

Regarding memory usage, we observe that \tool{} consumes more memory across all datasets. However, even on the most complex datasets—c, cpp, java, lua, and rust, the memory usage remains under 5GB, which is acceptable. The increased memory consumption mainly results from \tool{}’s stronger generalization
and its use of the L* algorithm for lexical inference, which increases the number of oracle queries and necessitates caching their results.

\begin{framed}
\textbf{Answer to RQ2:}
\tool{} demonstrates strong scalability on complex real-world languages (i.e., C, C++, Java, Lua, Rust) successfully inferring grammars with average runtime of 13.6 hours, where SOTA methods fail to complete within 48 hours. On smaller datasets, 
\tool{} accelerates average running time by 1.56$\times$ and 1.39$\times$ over \sotat{} and \sotak{}, respectively, while requiring 1.68$\times$ and 1.99$\times$ more oracle calls due to its more precise lexical inference and the stronger generalization enabled by decomposition forests. 
\end{framed}

\subsection{RQ3: Grammar Readability}
Readability is essential in contexts such as program understanding, where grammars must be easily interpretable by humans. Another important use case concerns semantic exploration: overly complex grammars often contain numerous ambiguities, which can lead to difficulties in downstream analysis. In this RQ, we evaluate the readability of inferred grammars using the following metrics:
\textbf{Grammar Size}: We calculate the count of unique non-terminals and terminals, along with the number of 
    productions and the length of each production (\textit{i.e.}, the length of the right-hand-side sequence of terminals and nonterminals). The size of the grammar is the sum of the lengths of all productions.
\textbf{Parse Time \& Memory}: We assess the average time per program and the peak memory usage required to parse the test programs using a parser generated from a grammar inferred by \sotak{} or \tool{}.

It is also important to note that \tool{} does not use context-free grammars to describe lexical rules. Instead, lexical analysis is handled separately. In contrast, \sotat{} represents both lexical and grammar rules uniformly using CFGs. Consequently, \tool{} cannot be directly compared to \sotat{} in terms of overall grammar size or structure. Therefore, the comparison in Table~\ref{tab:Readbility} focuses exclusively on the grammar, where both \tool{} and \sotak{} operate under the same CFG-based assumption.

Table~\ref{tab:Readbility} reports the grammar sizes and parsing performance for grammars inferred by \sotak{} and \tool{}. For small and medium-scale datasets (i.e., arith, math, fol, json, lisp, turtle, while, xml, curl , tinyc, minic), the grammars produced by the two SOTA tools are comparable in terms of the number of non-terminals (NT), the number of rules (A), average rule length, and total rule size. On mathexpr and tinyc, \tool{} produces grammars with more non-terminals, while on arith and turtle it yields fewer NTs. These differences reflect variations in grammar construction strategies, and note that \tool{} does not perform grammar simplification.

We also observed that parsing efficiency exhibits a mixed pattern across datasets. Specifically, \tool{} achieves lower parsing time and memory usage on arith, json, and lisp, while parsing time and peak memory consumption are higher on fol, while, and minic. For large-scale languages, parsing remains computationally expensive but feasible; for example, parsing a single Java program requires 8.47 seconds and 3.06 GB of peak memory.

The primary advantage of \tool{} lies in its ability to scale to large, real-world programming languages and Rust, where \sotak{} fails to produce any grammar. This scalability substantially extends the practical applicability of grammar inference.

\begin{table}
\setlength{\abovecaptionskip}{2pt} 
\setlength{\belowcaptionskip}{2pt} 
\footnotesize
\caption{Grammar size and parse performance on R0: Averages for grammars inferred in 10 runs; NT/T = unique (non-) terminals; A = rules (alternatives); l(A) = avg. rule length; S = sum of rule lengths; $t_P$ = time to parse one test program (ks); $m_P$ = peak memory while parsing (GB); bold = \tool{} better than \sotak{}.}\label{tab:Readbility}
\setlength{\tabcolsep}{3pt}
\begin{tabular}{c|ccccccc|ccccccc}
  \hline
  \multicolumn{1}{c|}{}& \multicolumn{7}{c|}{\sotak{}}  & \multicolumn{7}{c}{\tool{}} 
   \\     &   NT & A &	l(A)	&S	& T &	$t_P$ &	$m_P$ & NT & A &	l(A)	&S	& T&	$t_P$ &	$m_P$  \\\hline
arith & 1 & 3 & 2.3 & 7 & 4 & .06 & .22 & 2 & 4 & 2 & 8 & 4 & \textbf{.02} & \textbf{.19} \\
math & 7 & 18 & 2.1 & 39 & 11 & .09 & .12 & 13 & 36 & 1.7 & 61 & 15 & \textbf{.04} & .16 \\
fol & 4 & 16 & 3.8 & 60 & 18 & .02 & .11 & 5 & 18 & 3.9 & 70 & 19 & .04 & .27 \\\hline
json & 5 & 17 & 2.1 & 36 & 13 & 0 & .04 & 6 & 18 & 2 & \textbf{35} & 13 & .01 & .07 \\
lisp & 2 & 5 & 3 & 15 & 5 & .03 & .14 & 4 & 9 & 2.3 & 21 & 6 & \textbf{.01} & .17 \\
turtle & 3 & 7 & 3.3 & 23 & 9 & 0 & .06 & \textbf{2} & \textbf{7} & 4 & 28 & 9 & .01 & .08 \\
while & 5 & 14 & 3.8 & 52 & 17 & .01 & .18 & \textbf{5} & \textbf{14} & 3.7 & \textbf{51} & 17 & .05 & .34 \\
xml & 5 & 17 & 3.5 & 59 & 12 & .01 & .05 & \textbf{5} & \textbf{17} & 2.8 & \textbf{48} & 12 & \textbf{.01} & .08 \\\hline
curl & 15 & 24 & 1.6 & 39 & 8 & .03 & .03 & \textbf{10} & 26 & 1.6 & 42 & 10 & \textbf{0} & .1 \\
tinyc & 3 & 15 & 3.1 & 47 & 14 & .02 & .12 & 6 & 19 & 2.8 & 52 & 14 & \textbf{.02} & \textbf{.11} \\
minic & 12 & 30 & 2.7 & 80 & 17 & .01 & .04 & \textbf{11} & 34 & 2.7 & 91 & 17 & .06 & .58 \\\hline
c & - & - & - & - & - & - & - & \textbf{46} & \textbf{147} & 1.9 & \textbf{286} & 35 & \textbf{4.31} & \textbf{1.39} \\
cpp & - & - & - & - & - & - & - & \textbf{77} & \textbf{248} & 1.9 & \textbf{461} & 48 & \textbf{14.86} & \textbf{2.44} \\
java & - & - & - & - & - & - & - & \textbf{89} & \textbf{280} & 2.2 & \textbf{610} & 53 & \textbf{8.47} & \textbf{3.06} \\
lua & - & - & - & - & - & - & - & \textbf{34} & \textbf{122} & 1.7 & \textbf{212} & 31 & \textbf{10.18} & \textbf{1.94} \\
rust & - & - & - & - & - & - & - &\textbf{ 75} & \textbf{260} & 1.9 & \textbf{490} & 56 & \textbf{14.59} & \textbf{2.19} \\

\hline

\end{tabular}

\end{table}

\begin{framed}
\textbf{Answer to RQ3:}
Compared to \sotak{}, \tool{} produces grammars with 1.11$\times$ the number of non-terminals, 1.21$\times$ the number of rules, and 1.11$\times$ the total rule length, indicating similar overall complexity and readability. Notably, these results demonstrate that despite its stronger generalization capability and scalability, \tool{} maintains comparable grammar simplicity.
\end{framed}

\section{Discussion and Future Work}

\textbf{Effectiveness of Each Module.} From a theoretical perspective, the distributional matrix based inference is the most influential component. It introduces Deterministic Limited Correctness Verification rather than relying on uncertain and incomplete sampling, as the existing work \sotak{} does, to ensure absolute correctness of the grammar, allowing bounded correctness guarantees while exploiting distributional information to construct concise grammars with strong generalization. Compared to \sotak{}, it only requires enumerating maximal cliques (rather than all subsets), making traversal computationally feasible in practice. On the other hand, the decomposition forest reduces the problem scale, enabling the computation of more information for grammar construction, thereby ensuring computational feasibility. It enables repeated decomposition of a sample until the grammar fully incorporates all rules present in that sample, thereby preventing the rule-loss problem observed in \sotak{}, and can also be applied to lexical inference for significant efficiency gains. While both components are essential, the distribution-matrix-based inference provides the core advancement driving \tool{}’s superiority.

\textbf{Improving parsing efficiency}. The grammars inferred by \tool{} for complex languages exhibit low parsing efficiency: parsing a single input may take several seconds or even over ten seconds. In contrast, parsing inputs from simpler languages typically takes only tens of milliseconds.
The inefficiency mainly arises from two factors: (1) the inherent complexity of the input itself, and (2) ambiguity in the inferred grammar. While input complexity is unavoidable, future improvements can be made by refining the way bubbles are merged during generalization on the parse tree to reduce grammar ambiguity and improve parsing efficiency.

\textbf{More accurate grammar}.
\tool{} ensures first-order swap correctness on the parse trees built during inference. However, this does not imply that the inferred grammar is entirely correct. On the one hand, the distributional matrix captures only first-order swap information and fails to reflect higher-order structures, whose computation is infeasible due to high complexity. On the other hand, although the parse trees built during generalization enable first-order swap correctness for the observed samples, grammar ambiguity may cause different parse trees to be produced during actual parsing, making it difficult to guarantee the correctness of the swap operations. Furthermore, since a grammar is a compressed representation of a potentially infinite set of strings, it is inherently difficult to perform complete verification. Overall, there exists a trade-off between generalization and precision in grammar inference: while \tool{} demonstrates stronger generalization than existing methods, it may come at the cost of slightly reduced precision.

\section{Conclusions}

We present \tool{}, a novel grammar inference method that effectively addresses the scalability challenges of inferring grammars for complex programming languages such as C, C++, and Java. By leveraging a decomposition forest and a distributional matrix, \tool{} significantly outperforms existing state-of-the-art methods — including \sotaa{}, \sotat{}, and \sotak{} — which fail to produce results within practical time limits on these languages. While parsing efficiency for the complex programming languages remains an open challenge, \tool{} represents a substantial advancement by generating grammars that are both effective and human-readable.

\begin{acks}
This work is supported by the Natural Science Foundation of Guangdong Province (2025A1515011946), and by the Major Key Project of PCL (PCL2023A06-4).
\end{acks}
\bibliographystyle{ACM-Reference-Format}
\bibliography{sample-base}

\appendix

\end{document}